# *Research on Mode Transition of Micro-Newton Cusped Field Hall Thruster

WU Jiahao [1)#] ZENG Ming [1†)] LIU Hui [1)#] YU Daren [1†)]

1)（Harbin Institute of Technology，School of Energy Science and Engineering，Key Laboratory of Aerospace Plasma Propulsion，Harbin 150001）

## Abstract

The micro-newton-level cusped field Hall thruster is an electric propulsion device that employs microwave-assisted ionization control. It functions as an actuator in drag-free control systems, ensuring control accuracy and stability through continuously adjustable thrust across a wide range. However, a mode transition during regulation can induce abrupt changes in anode current, thereby compromising control precision and stability. Consequently, it is essential to investigate the underlying mechanisms of this mode transition. This study examines variations in internal plasma parameters and discharge characteristics before and after the microwave-induced mode transition, primarily using probe diagnostics. Experimental results show that prior to the transition, the plasma luminous region is concentrated mainly within the electron cyclotron resonance (ECR) zone, located approximately 1-3 mm upstream of the anode. Following the transition, the luminous region shifts further upstream, and the plasma density near the anode exceeds the cutoff density, exhibiting a sharp axial decline. This density variation alters the propagation characteristics of fundamental waves, thereby changing the electron heating mechanism. When plasma density reaches the cutoff density, the R-wave and O-wave-responsible for driving

---

ionization-are rapidly attenuated or reflected. Under these conditions, the R-wave fails to reach the resonance layer, rendering ECR-dominated ionization ineffective. The ionization mechanism consequently shifts from R-wave and O-wave dominance to O-wave dominance alone. Accordingly, electron heating transitions from volume heating to surface wave heating. These findings provide a foundation for optimizing microwave transmission within the thruster and for lowering the threshold at which mode transition occurs.

# 1 Introduction

In 2015, the Laser Interferometer Gravitational-Wave Observatory (LIGO) in the United States made the first detection of gravitational waves, formally inaugurating the era of gravitational-wave astronomy. To detect low-frequency gravitational waves, Japan proposed the DECi-hertz Interferometer Gravitational-wave Observatory (DECIGO) mission[1], while Europe and the United States proposed the Laser Interferometer Space Antenna (LISA) mission[2] and launched the LISA Pathfinder technology demonstration satellite in 2015[3]. In February 2024, the European Space Agency (ESA) and the National Aeronautics and Space Administration (NASA) officially approved the LISA mission, with a planned launch of the space-based gravitational-wave observatory in 2035[4,5]. Furthermore, China's *National Medium- and Long-Term Development Plan for Space Science (2024-2050)*, released in October 2024, identifies space-based gravitational-wave detection as one of the major candidate missions expected to yield landmark scientific breakthroughs in the next phase. Gravitational-wave detection satellites rely on drag-free control systems to ensure stable operation of the test masses. As the actuator of the drag-free control system, micro-thrusters provide continuous thrust adjustment over a wide range thereby maintaining the control accuracy and stability of the system[6].

The micro-Newton cusped-field Hall thruster represents a key micro-propulsion technologyunder China's National Key R&D Program the "Gravitational-Wave Special Project," and its performance requirements include wide-range, continuously adjustable thrust[7-9]. The thruster has already achieved a thrust regulation range of 1-100 μN, demonstrating strong engineering potential applicability[10]. Its fundamental operating principle involves generating plasma through microwave electron cyclotron resonance (ECR) discharge within a cusped magnetic field and accelerating ions by means of an anode electric field to produce thrust[11-13].

To achieve high-precision drag-free control, the output thrust of the micro-Newton cusped-field Hall thruster must exhibit a monotonic and continuous relationship with its input control parameters. However, experimental studies involving continuous adjustment of operating parameters have revealed a mode transition phenomenon, as illustrated in Fig. 1. Specifically, when only a single input parameter—such as anode voltage or microwave power-is varied, key performance parameters like thrust or discharge current do not change smoothly; instead, they exhibit abrupt and discontinuous jumps. This unexpected discontinuity directly disrupts the ultraquiet stability required for drag-free control. Moreover, the accompanying hysteresis loop causes the transition points to differ depending on whether the parameter is increased or decreased, thereby impairing the drag-free control system's ability to precisely track and compensate for thrust variations. Consequently, both the accuracy and stability of drag-free control are severely compromised.

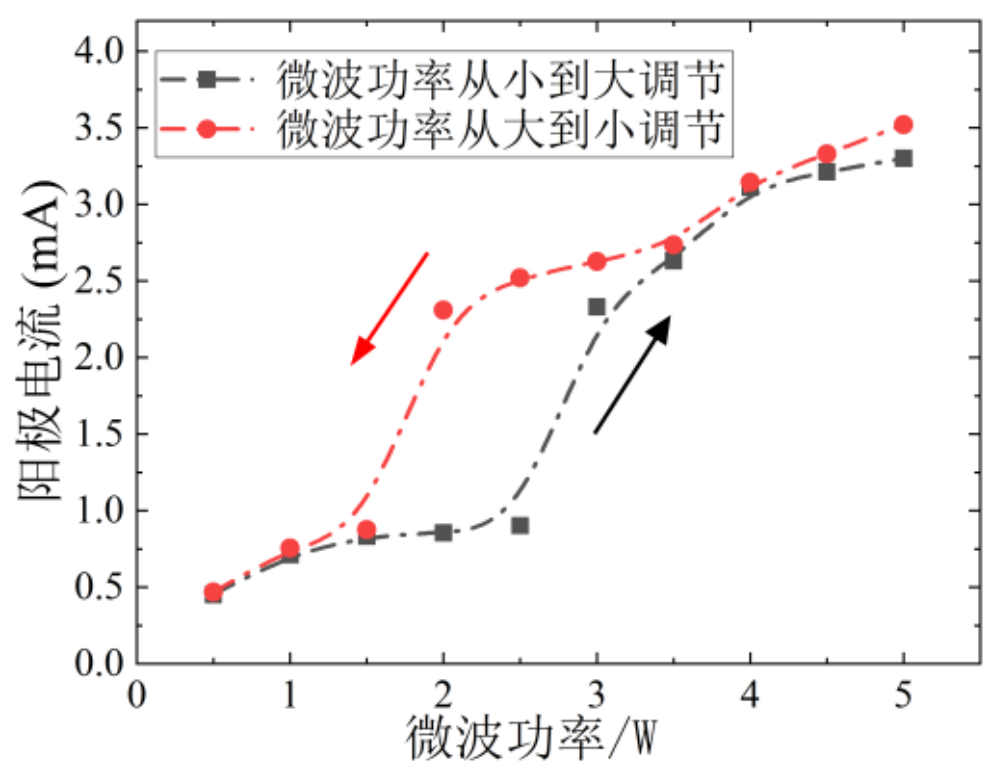

Fig. 1. Mode transition during microwave regulation.

Researchers from Kyushu University and other institutions in Japan were among the first to observe mode transition phenomena[14]. In experiments on$\mu_1$ and $\mu_{10}$ plasma thrusters[14-20], they found that variations in microwave power or propellant flow rate induced abrupt jumps in screen-grid current. By comparing four distinct waveguide configurations[18], they demonstrated that waveguide geometry significantly alters the microwave propagation characteristics within the chamber, thereby influencing the location of plasma generation and the transport efficiency. Differences in waveguide structure lead to variations in energy coupling efficiency; when power is adjusted, these differences trigger transitions between two distinct operating modes.Further investigation revealed[19] that such mode transitions fundamentally arise from changes in the thruster's electron-heating mechanism.

Building on this foundation, Ding's team at Dalian University of Technology[21-26] conducted in-depth mechanistic studies. Through measurements of microwave reflection coefficients, optical emission spectroscopy, and theoretical modeling, they found that the internal magnetic field distribution and dynamic variations in microwave power affect plasma generation. During microwave power adjustment, power dynamically redistributes between absorption and reflection by the plasma load. This redistribution, in turn, modulates the plasma ionization rate and density. They proposed that during parameter tuning, changes in plasma density alter the microwave-plasma energy-coupling state, and that this dynamic instability ultimately drives mode transitions[26].

Meanwhile, Zeng's team at Harbin Institute of Technology[27] offered another important perspective. They developed a simplified magnet-free microwave ion thruster and observed mode transitions — identified via screen-grid current variations-during microwave power modulation. Langmuir probe measurements combined with microwave reflection-coefficient (Γ) data revealed a strong correlation between mode transition and the microwave cutoff density. When the plasma density exceeds the microwave cutoff density ($n_e > n_c$), the heating mechanism shifts from volumetric heating to surface-wave heating. This transition directly causes abrupt changes in thrust and discharge current.

Collectively, these studies—focusing on waveguide geometry, microwave energy coupling, and cutoff density-demonstrate that microwave propagation characteristics, electron-heating mechanisms, and plasma-density dynamics are key drivers of mode transitions in microwave plasma thrusters. These findings suggests that wave-plasma interactions and the evolution of plasma parameter space are central to understanding such transitions. However, current research lacks direct experimental evidence of how spatial plasma evolution within the thruster influences microwave energy propagation and absorption.

Microwave discharge is essential for the stable operation of micro-Newton cusped-field Hall thrusters and is critical for achieving wide-range continuous thrust control. Therefore, a deeper investigation into wave-plasma interaction mechanisms and plasma parameter space evolution in these thrusters is of great significance for elucidating the physics of mode transition and enhancing continuous thrust tunability. Accordingly, this study employs transmission-line parameter measurements combined with probe-based plasma diagnostics to investigate the characteristics of mode transition in micro-Newton cusped-field Hall thrusters, optimize thruster design, and advance the understanding of wave-plasma interactions and plasma-parameter evolution in microwave-discharge plasma thrusters.

# 2 Experimental Method

## 2.1 Experimental Approach

Mode transitions in micro-Newton cusped-field Hall thrusters are likely associated with wave-plasma interactions and the evolution of plasma parameter space. Therefore, intrusive probe diagnostics and discharge-parameter monitoring can effectively characterize the feature of different discharge modes and reveal the underlying transition mechanisms.

First, we identify the threshold operating conditions for mode transitions by monitoring the anode current and the standing wave ratio (SWR)/reflection coefficient, together with discharge imaging. The anode current is a key parameter for thrust control, while the SWR reflects the microwave coupling characteristics; together, these parameters provide reliable indicators of discharge state changes. We measure anode current and SWR over anode voltages ranging from 300 to 700 V and microwave powers from 1 to 4 W. When both parameters exhibit significant simultaneous changes, as corroborated by the discharge images.A mode transition is confirmed.

Next, we analyze the spatial evolution of plasma parameters under representative operating conditions. After establishing the transition thresholds, we measure ion current density in the exhaust plume using a Faraday probe, taking the center of the thruster exit plane as the originand a 15-cm scan arm, to characterize plume properties indifferent operating modes. Subsequently, we mapthe evolution of internal plasma parameters: setting the anode end face as $X$ = 0 mm and defining the outward channel direction as positive, a Langmuir probe mounted on a stepper motor acquires I-V curves from $X$ = -1 mm to $X$ = 4 mm in 1 mm increments. The electron temperature and plasma density are derived from these curves. Finally, we synthesize all data to elucidate the dynamics of mode transition.

## 2.2 Experimental Setup and System

As shown in Fig. 2(a), the micro-Newton cusped-field Hall thruster consists of permanent magnets, a boron nitride discharge channel, an anode, a microwave transmission line, and a resonant cavity. Gas injection orifices are arranged radially on the anode end face. The boron nitride ceramic channel is16 mm in length and has an inner diameter of 6 mm . A cusped magnetic field is generated by 8-mm and 4-mm samarium-cobalt permanent magnets arranged with like poles facing each other across a magnetic yoke; the resulting magnetic-field distribution is shown in Fig. 2(b). For the 2.45 GHz microwave source used in this study, the corresponding electron cyclotron resonance magnetic-field strength is 0.0875 T.

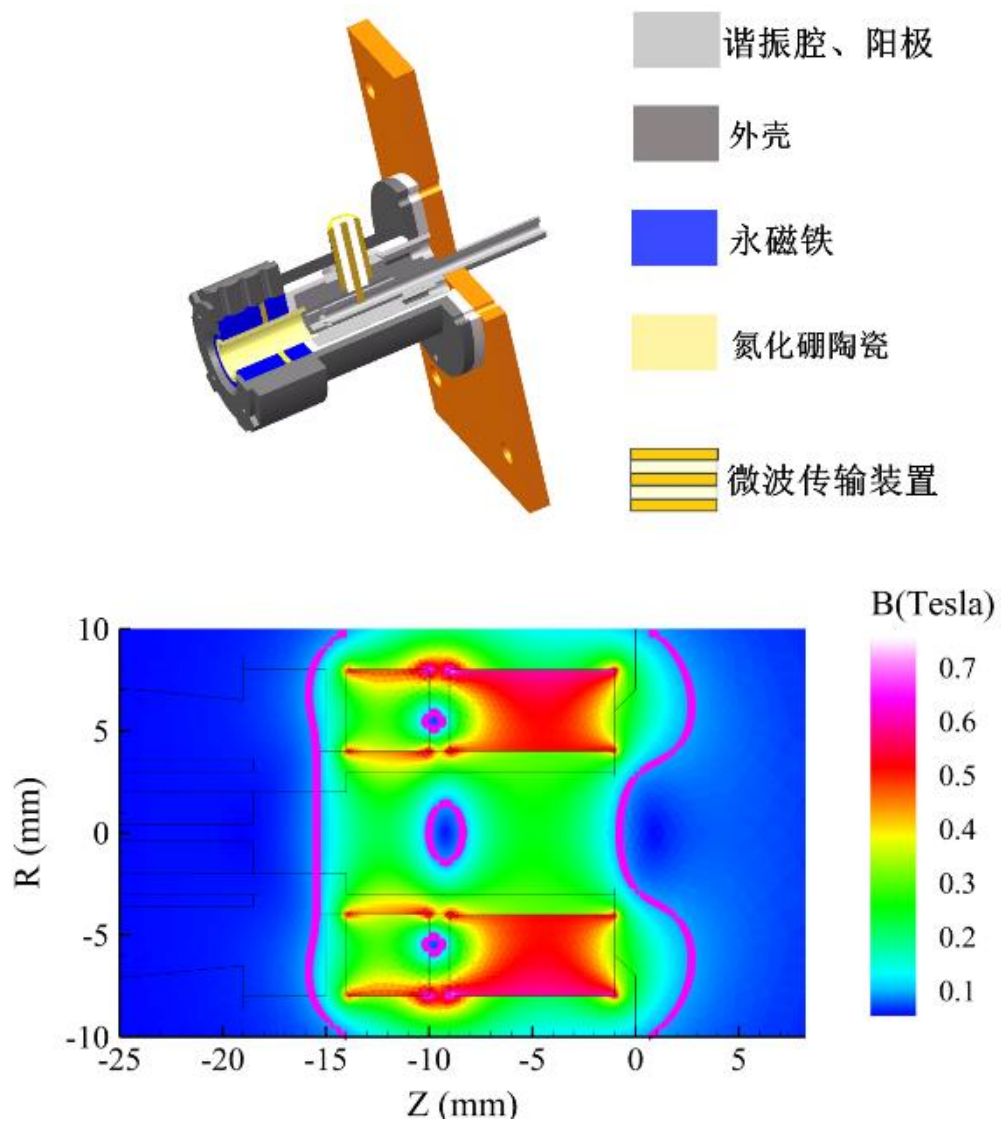

Fig. 2 Schematic of the thruster: (a) Structure; (b) magnetic field distribution.

The experimental platform features a vacuum chamber measuring 80 cm × 80 cm × 80 cm, capable of maintaining a vacuum level of $10^{-4}$ Pa during extended operation. The propellant flowmeter is an Alicat device with a measurement range of 0.01-0.5 sccm (1 sccm = 1 mL/min under standard atmospheric conditions). The power supply used in the experiment is a commercial  Earthworm Electronics (Shanghai) unit. The anode power supply provides a voltage range of 0 to 1000 V, the tungsten filament cathode keeper electrode power supply operates within 0 to 150 V, and the cathode heating power supply delivers a current range of 0 to 3 A. The microwave source operates at a frequency of 2.45 GHz and includes an internal system for measuring reflected power and standing wave ratio at both the transmission-line end and the thruster-load end. No tunable impedance-matching network is installed between the thruster and the microwave transmission system, and the transmission line employs a coaxial cable with a characteristic impedance of 50 Ω.

The Faraday probe and Langmuir probe measurement system used in the experiment are shown in Figure 3. A bias voltage is applied to the Faraday probe of -30 V to suppress electron interference. A measurement resistor of 198 kΩ is employed, and a capacitor is connected in parallel with this resistor to further suppress interference currents. In the Langmuir probe circuit, a bias voltage of -50 V is applied, the sweep voltage ranges from -50 V to 150 V, and the measurement resistor is 1 kΩ.

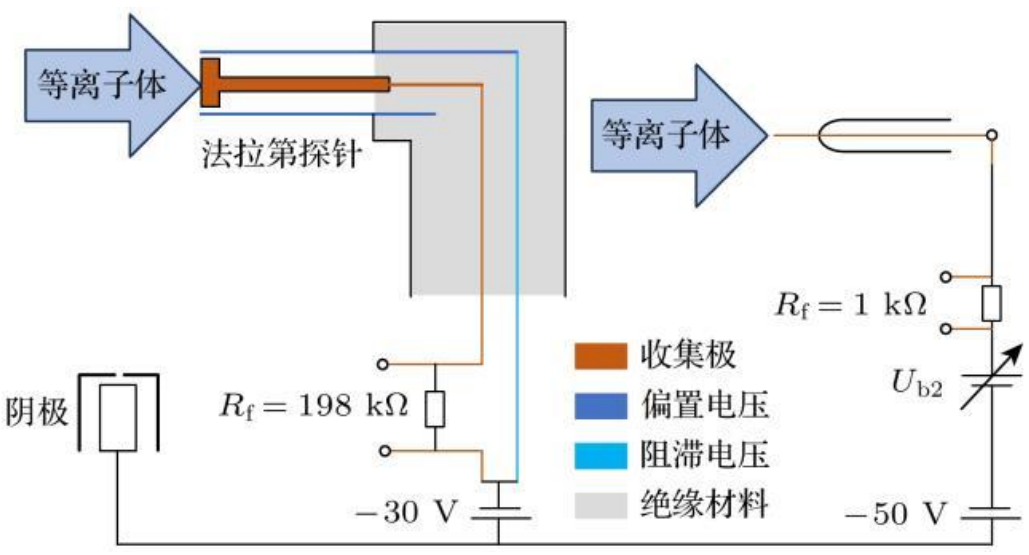

Fig. 3 Schematic of the Faraday probe and Langmuir probe measurement system.

Following the Langmuir probe measurements of the thruster-plasma *I-V* characteristics, we applied the floating-potential method[28] to correct the measured plasma density $n_i$. During plasma parameter calculation, the ion saturation current $I_{sat}$ increases with increasingbias voltage due to sheath effects. Consequently, we performed a linear fit to the measured current over the initial small voltage interval and adopted the fitted current value at the floating potential $\Phi_f$ as the ion saturation current $I_{sat}$. Conventional analyses often use Equation (1) to compute plasma density; however, the electron saturation current $I_{es}$ is typically difficult to determine. To improve the accuracy of plasma-density measurements obtained with the Langmuir probe , we calculated $n_i$ using the Bohm current and sheath model, as shown in Equation (2).

$$n_i = n_e = \frac{I_{es}}{eA}\sqrt{\frac{2\pi m_e}{kT_e}} \tag{1}$$

$$n_i = \frac{I_{sat}}{2\pi(R_p + d)eL\alpha_0 c_s} \tag{2}$$

Here, $I_{sat}$ denotes the saturated ion current; $R_p$ represents the probe radius; L is the probe measurement length; $\alpha_0$ is the Bohm current coefficient (experimentally calibrated to 0.6-0.7); $c_s = \sqrt{KT_e / M}$ is the ion sound speed; and d denotes the sheath thickness.

The sheath thickness is calculated using the following formula:

$$d = \frac{1}{3}\sqrt{\frac{2}{\alpha_0}}(2\eta_f)^{3/4}\lambda_D \approx 1.0\eta_f^{3/4}\lambda_D \tag{3}$$

Here, the normalized floating potential is defined as $\eta_f = |V_f / T_{eV}|$, and the Debye length is given by $\lambda_D = \sqrt{\epsilon_0 KT_e / ne^2}$ .

The Langmuir probe used in the measurements consisted of a tungsten wire with a radius of 1 mm and an exposed length of 2 mm. We initialized the plasma density as $n_0 = 10^{16}$ $m^3$. Substituting $n_0$ into the Debye-length formula yielded the initial value $\lambda_{D0}$ after which an iterative calculations were performed to obtain $n_i$. In practice, repeated measurements under the same operating condition were necessary because offactors such as probe contamination andbombardment by the thruster plume .

# 3 Results and Discussion

## 3.1 Mode Transition Phenomenon in Micro-Newton Cusped-Field Hall Thrusters

During the measurements, it was found thatwhen the microwave power was fixed at 4 W and the anode voltage was varied from 300 to 700 V , the anode current under the 500 V/0.4 sccm condition increased abruptly, as shown in Fig. 4(a). Subsequently, measurements were performed at a fixed anode voltage of 500 V while varying the microwave power from 1 to 5 W was measured. At 3 W/0.3 sccm and 3 W/0.4 sccm, the anode current decreased sharply, as shown in Figure 4(b). The results show that the coupling characteristics between the DC field and the microwave field inside the thruster and the variation of thepropellant flow rate affect the occurrence of mode transitions.

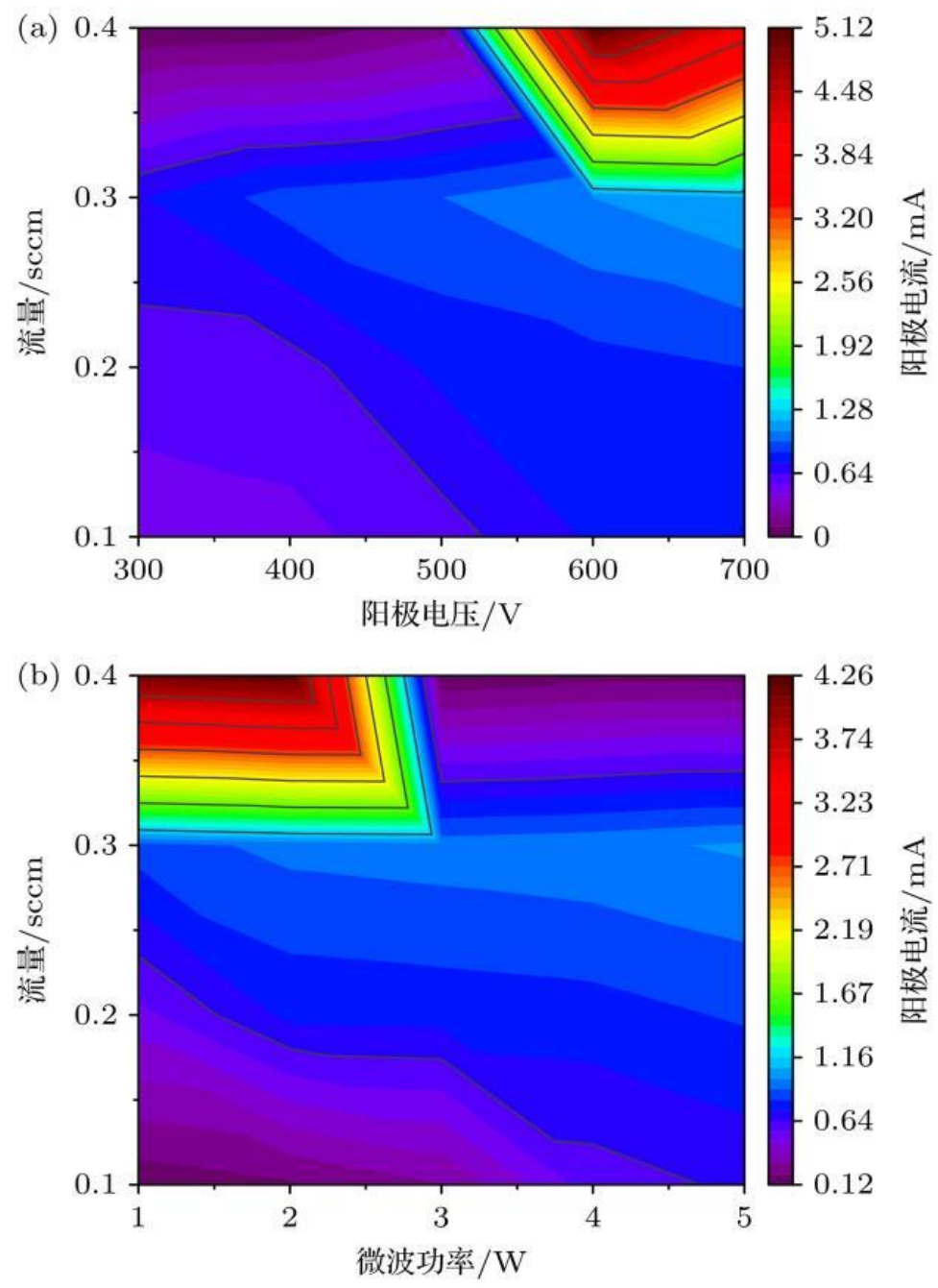

Fig. 4 Results of anode voltage and microwave power regulation: (a) Variation of anode current with anode voltage adjusted from 300 to 700 V at a constant microwave power of 4 W; (b) variation of anode current with microwave power adjusted from 1 to 5 W at a constant anode voltage of 500 V.

Discharge imaging of the thruster channel during operation revealed significant differences in the ECR ionization region between the two operating conditions-0.3 sccm/2 W and 0.4 sccm/4 W illustrated in Fig. 5. Under the 0.4 sccm/4 W condition (Fig. 5(b)), the plasma luminous region contracted noticeably toward the rear of the anode compared with the 0.3 sccm/2 W condition (Fig. 5(a)). This shift indicates a change in plasma discharge characteristics.

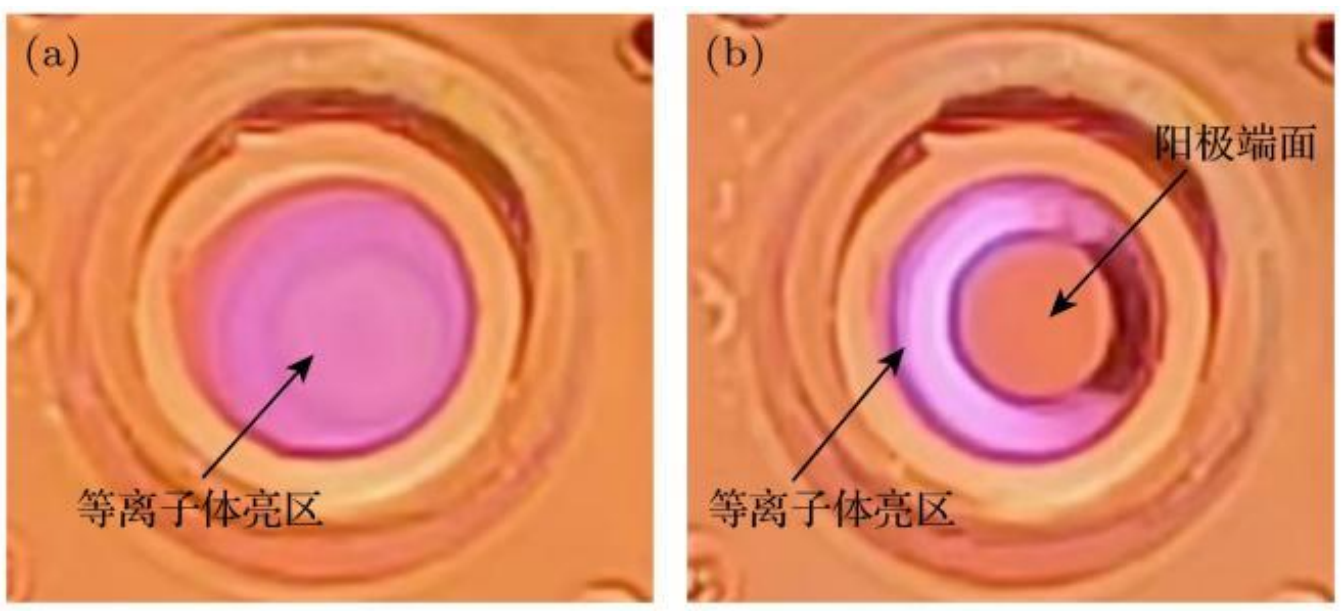

Fig. 5 Distribution of the plasma luminous region under two different operating conditions: (a) plasma luminous region upstream of the anode at 0.3 sccm and 2 W; (b) plasma luminous region receding toward the anode end-face at 0.4 sccm and 4 W.

To further investigate the mechanism underlying mode transitions, we measured the transmission-line parameters and anode-current responses during anode-voltage modulation under two representative operating conditions: 3 W/0.2 sccm and 4 W/0.4 sccm, as shown in Fig. 6. Under the 4 W/0.4 sccm condition, the standing wave ratio and reflection coefficient $\Gamma$ were 2 and 0.33, respectively; under the 3 W/0.2 sccm condition, the corresponding values were 1.2 and 0.1, respectively. Notably, neither parameter exhibited significant variation with increasing anode voltage. The anode current under the 3 W/0.2 sccm condition increased steadily with voltage, whereas under the 4 W/0.4 sccm condition, the current response rose sharply between 500 and600 V. This observation indicates that, in a DC-microwave-coupled cusped Hall thruster, DC voltage regulation influences plasma states without substantially altering microwave transmission matching characteristics. Consequently, variations in transmission line parameters primarily originate from the microwave control process.

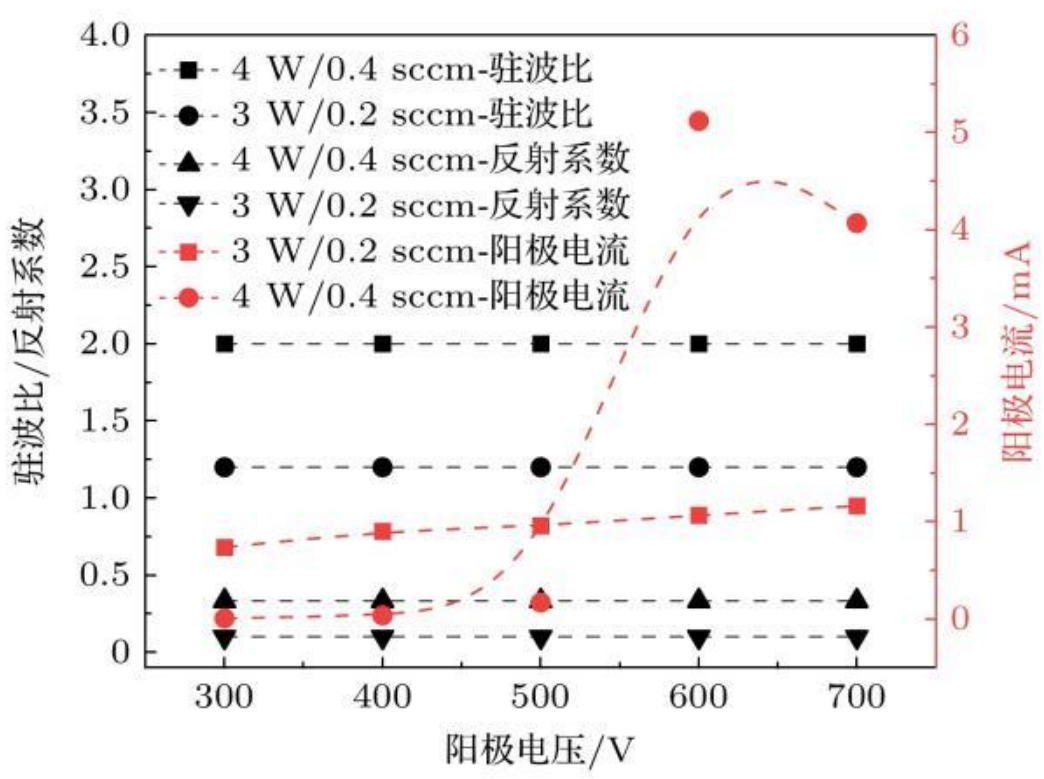

Fig. 6 Anode current and VSWR/reflection coefficient results under two representative operating conditions during voltage regulation.

Under zero-DC voltage conditions, we independently varied the microwave power (1-5 W, with a fixed flow rate of 0.3 sccm) and the propellant flow rate (0.1-0.5 sccm, with a fixed microwave power of 4 W). The corresponding responses of the reflection coefficient and voltage standing wave ratio (VSWR) are shown in Fig. 7. The experiments revealed that within the microwave power range of 2-3 W (Fig. 7(a)) and the propellant-flow-rate range of 0.3-0.4 sccm (Fig. 7(b)), the VSWR increased abruptly from 1.2 to 2 and from 1.3 to 2, respectively, while the reflection coefficient increased from 0.1 and 0.11 to 0.3. According to transmission-line theory, such abrupt changes in the reflection parameters indicate a significant change in load impedance. In principle, a VSWR of 2 corresponds to a microwave power transmission loss of approximately 11%. At an incident power of 4 W, the measured reflected power reached 1.3 W , further confirming a sudden change in plasma impedance during the mode transition.

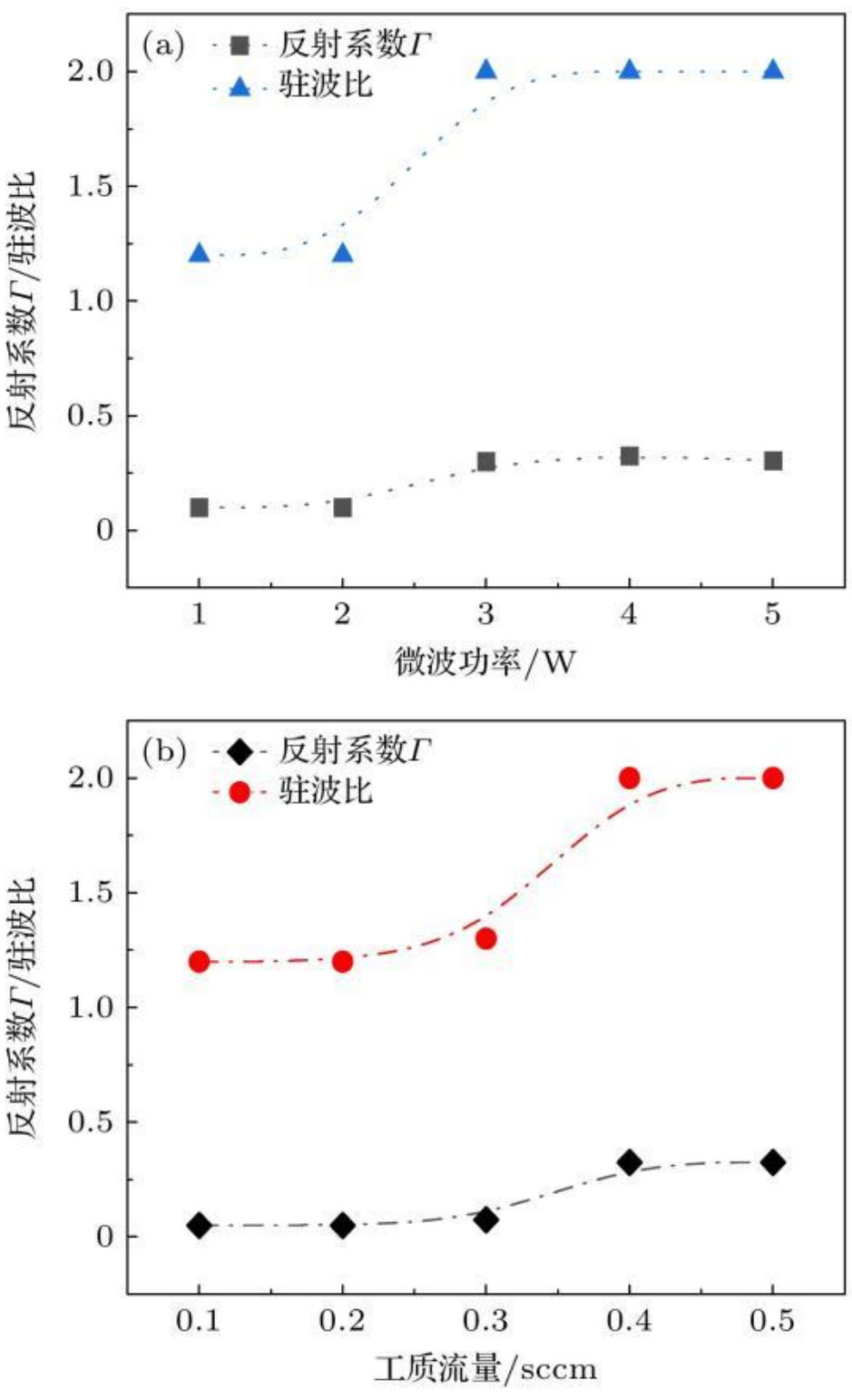

Fig. 7 Abrupt changes in VSWR and reflection coefficient with control parameters in the absence of a DC voltage: (a) Significant increase in VSWR and reflection coefficient as microwave power varies from 1 to 5 W; (b) sharp rise in VSWR and reflection coefficient as propellant flow rate increases from 0.1 to 0.5 sccm.

Collectively, these results indicate that mode transition in micro-Newton cusped-field Hall thrusters is primarily governed by microwave discharge and influenced by propellant flow rate. The main differencesbetween the operating modes before the transition (exemplified by 0.3 sccm/2 W) and after the transition (exemplified by 0.4 sccm/4 W) can be summarized as follows:

1) Location of the plasma luminous region: It contracts and shifts from the electron cyclotron resonance (ECR) zone near the anode tip (1-3 mm) to the vicinity of the anode face;

2) Microwave coupling efficiency: The voltage standing wave ratio (VSWR) increases markedly from approximately 1.2 to 2, accompanied by a substantial rise in microwave reflectance;

3) Discharge-currentbehavior: The anode current exhibits an abrupt change during parameter adjustment.

## 3.2 Spatial Evolution of Plasma Parameters During the Mode Transition: Measurement Results and Mechanistic Analysis

As discussed in Section 3.1, the microwave-driven mode transition induces displacement of the ECR ionization zone. Consequently, it is necessary to measure the spatial evolution of plasma density under microwave-dominated discharge conditions. Using a Faraday probe, we measured the ion current density in the thruster plume under 2 W/0.3 sccm (pre-transition) and 4 W/0.4 sccm (post-transition) microwave-dominated discharge conditions. The scanning results are presented in Fig. 8. Before the mode transition, the peak ion current density reached 0.073 A/m², exhibiting a unimodal plume profile. After the mode transition, the ion current density decreased significantly.

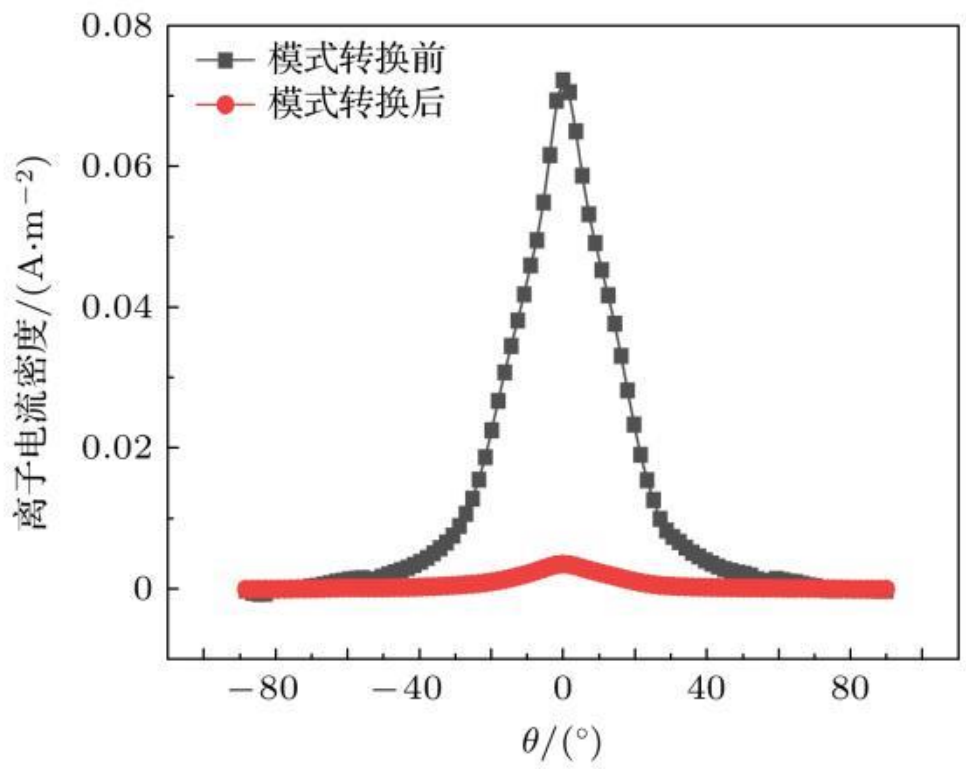

Fig. 8 Plume ion current distribution at 2 W/0.3 sccm (before mode transition) and 4 W/0.4 sccm (after mode transition).

Subsequently, we employed a Langmuir probe to measure the plasma density within the thruster channel before and after the mode transition. Fig. 9(a) and 9(b) present the *I-V* characteristics before and after the microwave-discharge mode transition, respectively. By differentiating the *I-V* curves, we obtained the first-derivative profiles shown in Fig. 10. The plasma space potential $\Phi_p$ was then determined from the extrema of these first-derivative curves. After smoothing the data, we selected the segment of each *I-V* curve lying between the floating potential $\Phi_f$ and the space potential $\Phi_p$. Taking the natural logarithm ln*I* of this segment, we performed linear regression, with the resulting fitted lines displayed in Fig. 11. Finally, the electron temperature $T_e$ was derived directly from the slope of these fitted lines.

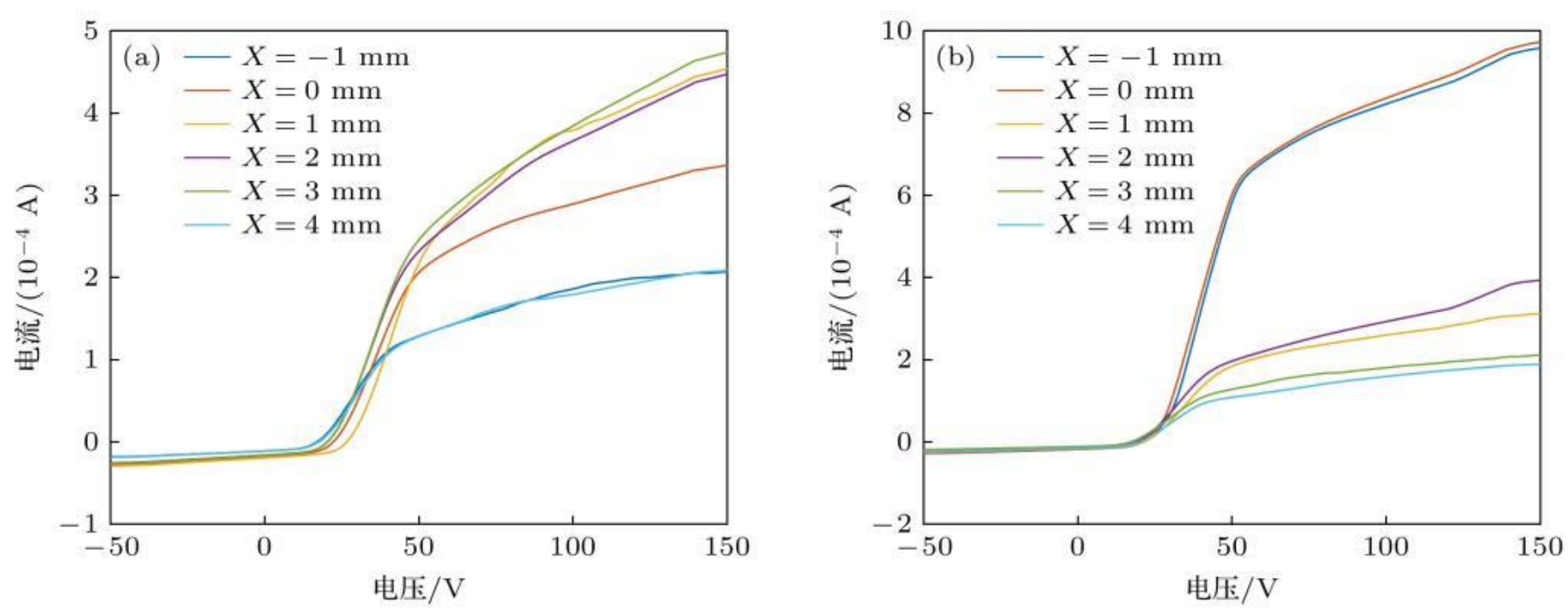

Fig. 9 *I-V* curves at the measurement points before and after mode transition: (a) *I-V* curves at positions from *X* = -1 to 4 mm before mode transition; (b) *I-V* curves at positions from *X* = -1 to 4 mm after mode transition.

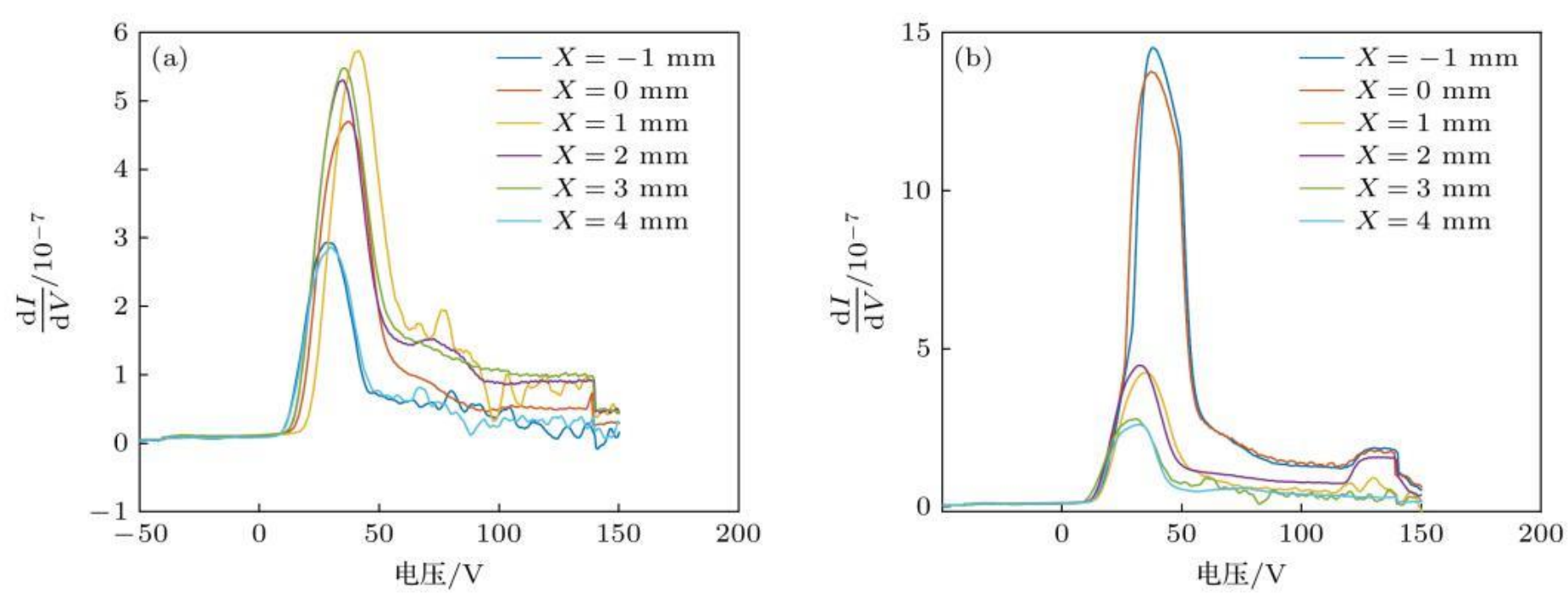

Fig. 10 Profiles of the first derivative of the *I-V* curves before and after mode transition: (a) Distribution of the first derivative for *I-V* curves at *X* = -1 to 4 mm before mode transition; (b) distribution of the first derivative for *I-V* curves at *X* = -1 to 4 mm after mode transition.

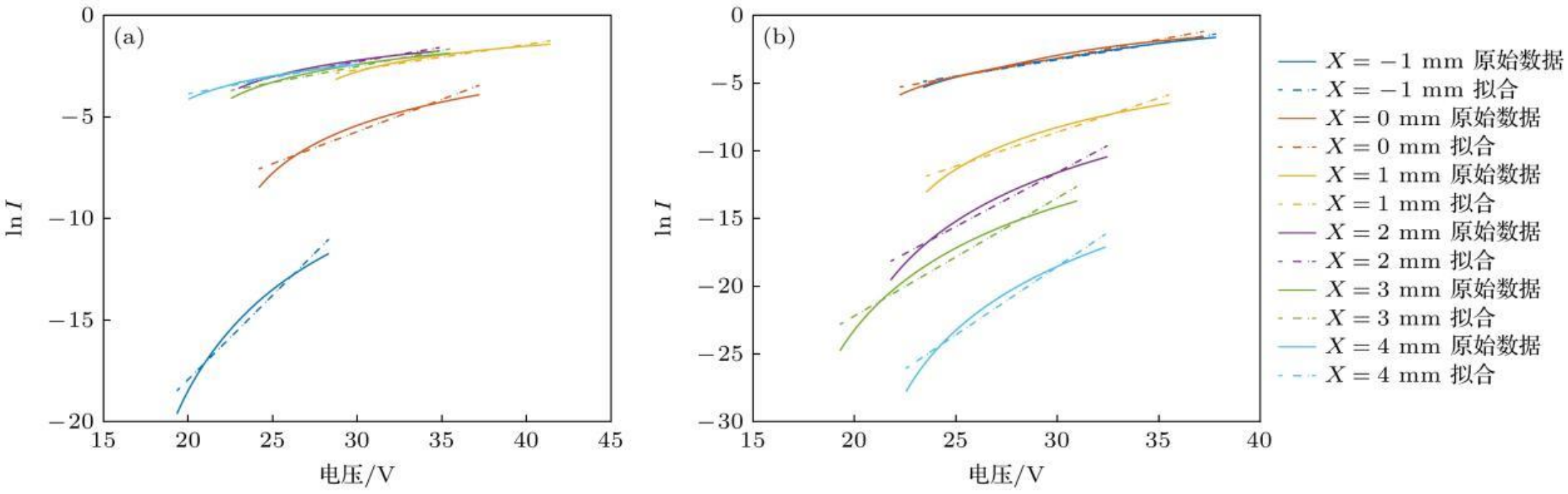

Fig. 11 Linear fitting results for the electron temperature before and after mode transition: (a) Linear fits to the electron temperature at measurement points from *X* = -1 to 4 mm before mode transition; (b) linear fits to the electron temperature at measurement points from *X* = -1 to 4 mm after mode transition.

After calculating $T_e$, we determined the plasma density $n_i$ using the floating-potential-corrected method. Fig. 12 presents the electron temperature and plasma density along the channel at various axial positions before and after mode transition. Prior to mode transition, as shown in Fig. 12(a), the electron density along the thruster axis is primarily concentrated near X = 1 mm from the anode end face. Similarly, the electron temperature peaks at approximately X = 1 mm from the origin along the axial direction, indicating that the primary ionization zone is localized within X = 1 mm upstream of the anode. Furthermore, according to the magnetic field distribution, this region coincides with the electron cyclotron resonance (ECR) surface within the thruster. This suggests that, under microwave-dominated discharge conditions prior to mode transition, plasma ionization is primarily driven by ECR heating. After mode transition, as depicted in Fig. 12(b), the plasma density is mainly confined to the region between X = -1 mm and X = 0 mm relative to the anode end face. Both plasma density and electron temperature progressively decrease toward the downstream direction. These results indicates that the ECR-dominated ionization mechanism has undergone a significant change within the thruster, and under this new dominant heating regime, plasma energy dissipates rapidly.

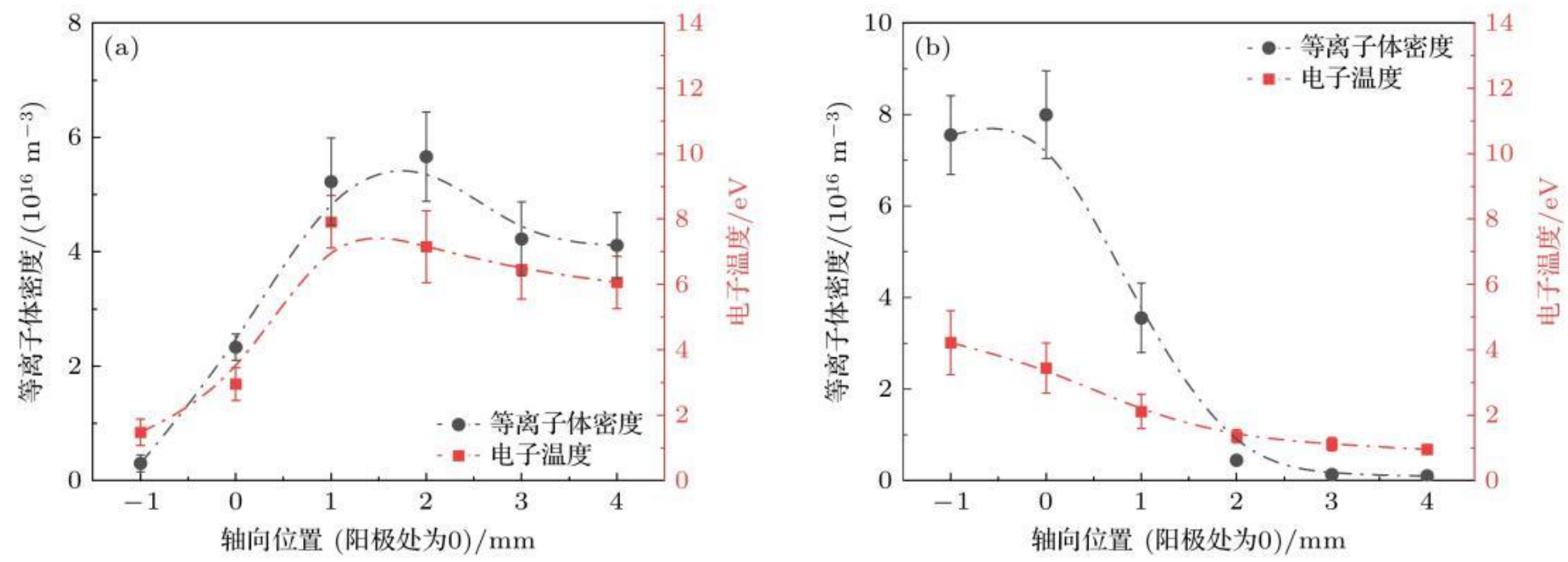

Fig. 12 Evolution of plasma parameters within the thruster channel during the mode transition process: (a) Electron temperature and plasma density at various measurement locations within the channel before mode transition; (b) electron temperature and plasma density at various measurement locations within the channel after mode transition.

Existing studies indicate that the primary factor driving changes in electron-heating modes in microwave-dominated discharges is the alteration of fundamental-wave propagation due to variations in plasma density[29]. In a cusped-field Hall thruster, the relative orientation between the microwave electric field and the magnetic field, together with the characteristic plasma density, determines that the two dominant fundamental waves responsible for ionization are the R-wave and the O-wave. Their propagation characteristics in a magnetic field can be represented on a CMA diagram[30,31], as shown in Fig. 13.

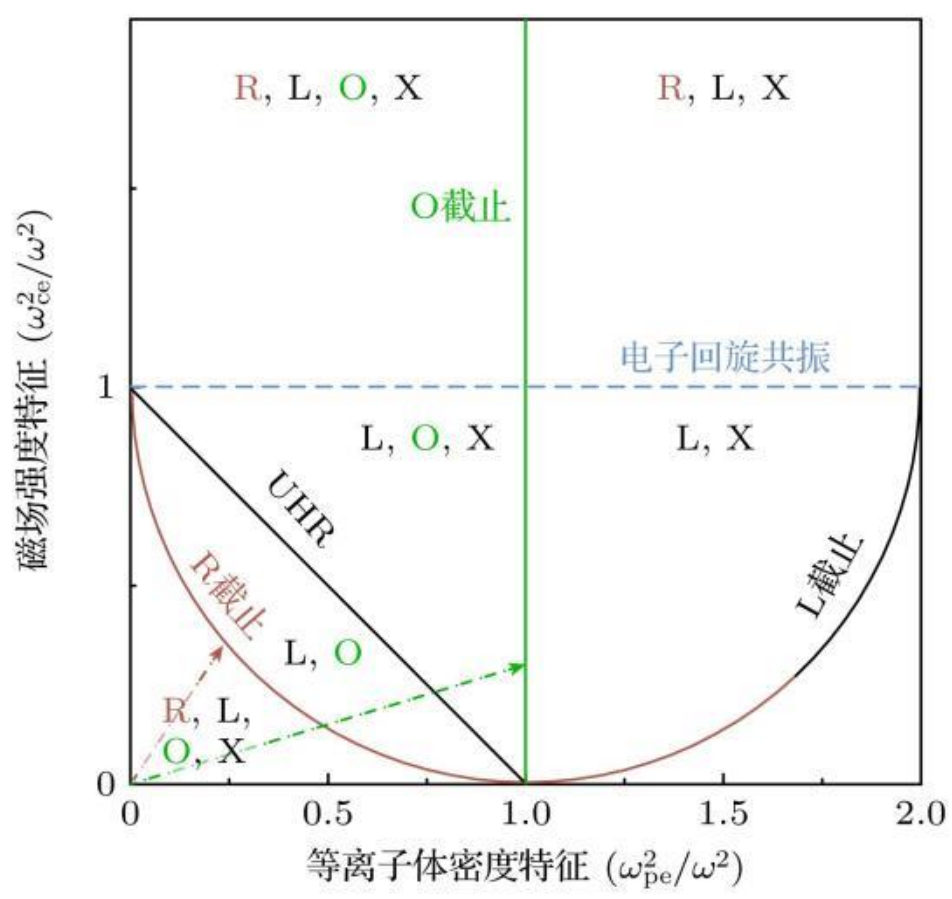

Fig. 13 Propagation characteristics of the R-wave and O-wave in magnetized plasma.

The microwave propagation path near the anode end face of the cusped-field Hall thruster extendsfrom a weak magnetic field region to a strong magnetic field region and from a low plasma density region to a high density region. The propagation paths of the R-wave and O-wave are illustrated in Fig. 13. The O-wave cutoff density is given by $n_{co} = \frac{m_e \epsilon_0}{e^2} \omega_{pe}^2$. At a microwave frequency $\omega$ = 2.45 GHz, $n_{co}$ = 7.45×10$^{16}$ m$^3$, with cutoff occurring where $\omega_{pe} = \omega$. The R-wave cutoff density is expressed as $n_{cR} = \frac{m_e \epsilon_0}{e^2} \left( \omega^2 - \omega \omega_{pe} \right)$. In the weak magnetic field region near the anode (where the characteristic magnetic field strength is below the ECR resonance field), the R-wave cutoff density is lower than $n_{co}$, causing cutoff at a more upstream axial location. As microwave power and propellant flow rate increase, neutral gas density rises, enhancing propellant ionization rate and electron collision frequency. Consequently, plasma density reaches the cutoff density, leading to rapid attenuation or reflection of the fundamental wave within one skin depth. This explains the increased reflectivity observed in Fig. 7(a). Under these conditions, the R-wave cannot reach the ECR resonance surface, resulting in reduced electron temperature and plasma density at the ECR resonance location ($X$ = 1 mm), as shown in Fig. 12(b). Because the R-wave-dominated ECR mechanism can no longer be sustained, electron heating near the anode shiftsfrom a regime of coexisting R-wave and O-wave heating to one dominated by the O-wave and surface waves.

When the plasma density near the anode reaches or exceeds the O-wave cutoff density, the transverse O-wave can no longer propagate through the plasma as a bulk wave. Consequently, the microwave energy-transport pathway changes: part of the O-wave undergoes skin-effect attenuation at the edge of the high-density plasma and deposits energy through ohmic heating, while another portion undergoes mode conversion at the plasma-ceramic interface, generating surface waves that propagate along the plasma boundary. This altered energy-transport process establishes an axially decaying plasma region within the thruster, gradually shifting the electron-heating mechanism from bulk heating to surface heating. Notably, this shift may cause the electron energy distribution to deviate from a Maxwellian profile; thus, a single electron temperature (Te) serves only as an approximation. Nevertheless, this does not affect the main conclusion regarding the fundamental transformation of the heating mode. The anode end face, which serves as the front end of the resonant cavity for microwave transmission, exhibits a design mismatch with the ECR resonance surface. Once the plasma density reaches the cutoff density, this mismatch causes microwave reflection in the near-anode region, thereby altering the plasma-heating mechanism within the thruster channel. Adjusting the magnetic-field configuration can expand the ECR resonance surface and thereby increase the probability of high-energy electrons being heated by the fundamental wave. In addition, optimizing the anode structure—for example, by adding sharp tips to the anode end face—can extend the microwave ionization zone, enabling microwave propagation downstream even after the plasma density exceeds the cutoff density at the ECR resonance surface.

## 4 Conclusions and Future Work

To address the demand for wide-range continuous thrust adjustment in space-based gravitational-wave detection, this study measured the discharge parameters and analyzed the discharge images of a micro-Newton cusped-field Hall thruster before and after the mode transition, yielding the following main conclusions.

1) Experimental results showthat during parameter tuning, the micro-Newton cusped-field Hall thruster undergoes a mode transition dominated by microwave discharge, the fundamental characteristic of which is a shift in the electron-heating mechanisms. At 2 W/0.3 sccm (pre-transition), the plasma luminous region remains stable near the resonance surface with high microwave coupling efficiency. In contrast, at 4 W/0.4 sccm (post-transition), the luminous region contracts, the reflectivity increases, and microwave coupling efficiency declines.

2) The essence of this mode transition lies in the plasma density exceeding the

microwave cutoff density, which causes the plasma to the reflect the incident microwave. This reflection terminates O-wave propagation and prevents the R-wave from reaching the ECR resonance surface, thereby shifting the dominant electron-heating mechanism from efficient ECR bulk heating to localized and less efficient surface-wave heating. As a result, the ionization zone contracts and energy dissipation intensifies, ultimately leading to severe degradation of thruster performance, as evidenced by the plume characteristics and discharge current.

Future work will focus on optimizing the magnetic-field topology to broaden the ECR resonance surface and on enhancing the downstream microwave transmission capability by incorporating sharp tips on the anode end face. These strategies are expected to lower the threshold for mode transition, improve microwave transmission efficiency, and suppress reflectivity under high-plasma-density conditions. In addition, more precise diagnostic techniques will be employed to further investigate the detailed evolution of the electron energy distribution function.